# MICROSCOPIC PEDESTRIAN SIMULATION MODEL TO EVALUATE "LANE-LIKE SEGREGATION" OF PEDESTRIAN CROSSING*

By Kardi TEKNOMO**, Yasushi TAKEYAMA*** and Hajime INAMURA****

## 1. INTRODUCTION

One of the objectives of the pedestrian analysis is to evaluate the effects of proposed policy on the pedestrian facilities before its implementation. The implementation of a policy without pedestrian analysis might lead to a very costly trial and error due to the implementation cost (i.e. user cost, construction time and cost, etc.). On the other hand, using good analysis tools, the trial and error of policy could be done in the analysis level. Once the analysis could prove a good performance, the implementation of the policy is straightforward.

The problem is how to evaluate the impact of the policy quantitatively toward the behavior of pedestrians before its implementation. Since the interaction of pedestrians cannot be well address using a macroscopic level of analysis, a microscopic level of analysis is the choice. However, the analytical solution of the microscopic pedestrian model is very difficult and simulation models are more practical approach[1].

A Microscopic Pedestrian Simulation Model (MPSM) is a computer simulation model of the pedestrian movement where every pedestrian in the model is treated as an individual. There are many types of MPSM and most of them do not relate with each other. Gipps[2] and Okazaki[3] have discussed the microscopic simulation using cost and benefit cell and the magnetic force model, respectively. Social force model was developed by Helbing[4], while Blue and Adler[5], developed a cellular automata model for pedestrians. The use of microscopic pedestrian simulation for evacuation purposes was developed by several authors[6)7)8]. They used a queuing network model for the MPSM.

The microscopic pedestrian simulation model developed in this paper is a physical based model similar to the social force or magnetic force model with forward and repulsion forces as the main force driver. The detail of the model, however, is somewhat different from those two models since it does not require target time as an input to the model. Exactly the opposite of the fact, the dissipation time is the output of the model, similar to the Evacuation models. Compared to other physical based models that use continuous approach, our simulation model, is a discrete event simulation without the queuing theory. Compared to the cellular automata model that uses much heuristic approach, physical based models contain more physical meanings than merely a computational advantage.

Though the simulation model has much advantage to evaluate some policies quantitatively, the model has some limitation as merely numerical experiment rather than complete real world simulation.

The next section describes in more detail about the development of the simulation model, section 3 discusses about the experiments of "lane-like segregation" policy on the crossing, followed with the results of the experiments in section 4.

## 2. MODEL DEVELOPMENT

Pedestrians in the microscopic simulation model are modeled as autonomous objects to be seen from above of the facilities. A pedestrian is modeled as a circle with a certain radius (uniform for all pedestrians). Each pedestrian has its own initial location, initial time, and initial velocity and predetermined target location (opposite to the initial location). These inputs can be determine by the user as a design experiment or specified randomly. A forward force drives pedestrian movements when there is no other pedestrian in the facility. The forward force makes the pedestrian path almost in a straight line.

The model uses a difference equation. A discrete model is used to avoid numerical integration due to a non-constant acceleration. The time $t$ represents the simulation clock that can be calibrated later into real time.

If the current location, velocity and acceleration are denoted by vector, $\mathbf{p}(t)$, $\mathbf{v}(t)$ and $\mathbf{a}(t)$, respectively, the basic dynamical model is given by

$$\mathbf{p}(t+1) = \mathbf{p}(t) + \mathbf{v}(t) \qquad (1)$$
$$\mathbf{a}(t) = \mathbf{v}(t+1) - \mathbf{v}(t) \qquad (2)$$

An intended velocity is equivalent to a vector that will direct the pedestrian from the current position into the target position. Since acceleration can be seen as the difference of velocities, it is equal to the difference between intended velocity and the current velocity. However, the acceleration is also corresponding to force (mass is a constant). Thus, the force is equivalent to the difference between the intended velocity and the current velocity.

If the target location is denoted by **e**, the forward force is given by

$$\mathbf{f}_f(t) = \frac{(\mathbf{e} - \mathbf{p}(t))\mu_{max}}{\alpha . d(t)} - \mathbf{v}(t) \quad (3)$$

The parameter $\alpha$ and $\mu_{max}$ are given as calibrated parameters, and $d$ is the distance between the current position to the target location. The first term on the right hand side is intended velocity. Factor $\mu_{max}/(\alpha . d)$ is a smoothing factor to reduce the fluctuation of the average speed.

When other pedestrians exist, the repulsive force, as the interaction with other pedestrians, is inputted to the autonomous system. The pedestrian is then optimizing the movement by taking the best path to go to the target location while avoiding other pedestrians. Two kinds of repulsive forces are working together with the forward force. One force is driving away the pedestrian actor while still quite far from other closest pedestrian, the other force strongly repulses against all other pedestrians in the surrounding.

The first repulsive force, $\mathbf{f}_a$, is working only if there is another pedestrian in front of the actor (within the sight distance). If there are many other pedestrians, it considers only the closest pedestrian to the actor. The force is directing the actor away from that closest pedestrian as shown in Figure 1.

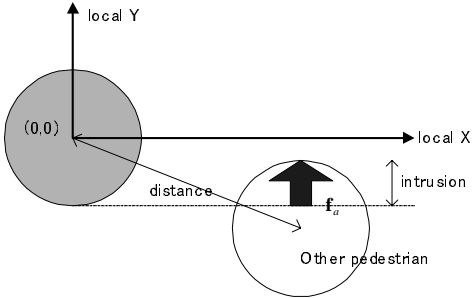

Figure 1. Force to repulse away

If $d$, $y$ and $r$ are, respectively, representing the distance between the pedestrians, intrusion of the closest pedestrian in the area in front of the actor and the influence radius of pedestrian, the first repulsive force in local coordinate is given by

$$\mathbf{f}_a(t) = \frac{\mu_{max}(2r - \mathbf{y}(t))}{\beta . d(t)} - \mathbf{v}(t) \quad (4)$$

The first term on the right hand side is the intended velocity and factor $\mu_{max}/(\beta . d)$ is the smoothing factor.

The local coordinate system is a translation and a rotation from the current coordinate system into the centroid of the actor. Since the angle of rotation depends on the current velocity, then the local X direction is the current velocity direction. If $\theta$ is the rotation angle, then the transformation from Euclidean coordinate system to local coordinate system is given by

$$\mathbf{l} = \mathbf{R}.(\mathbf{g} - \mathbf{p}) \quad (5)$$

Where $\mathbf{l}$ and $\mathbf{g}$ are local and Euclidean coordinates, respectively. The rotation matrix $\mathbf{R}$ is given by

$$\mathbf{R} = \begin{bmatrix} \cos\theta & \sin\theta \\ -\sin\theta & \cos\theta \end{bmatrix} \quad (6)$$

The first repulsive force in equation (2) should be transformed back to a current coordinate system using the inverse of the rotation matrix.

The second repulsive force, $\mathbf{f}_{i,r}$, is working to avoid the collision between pedestrians. It is generated when the influence radius of pedestrians overlap each other as shown in Figure 2. No repulsive force is generated if the influence radius does not overlap each other. Instead of considering the closest pedestrian as the first repulsive force, the second repulsive force considers all pedestrians surrounding and the forces are summed up linearly.

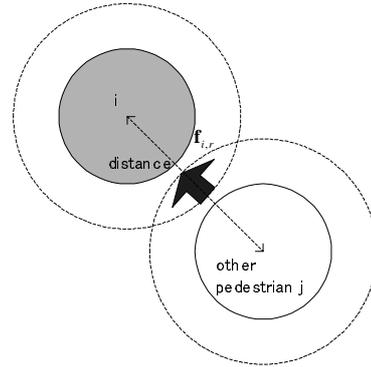

Figure 2. Repulsive force to avoid collision

Similar to the first repulsive force, the second repulsive force depends on the distance between the actor and other pedestrians surrounding it. The second repulsive force is given by

$$\mathbf{f}_{i,r}(t) = \frac{\mu_{max}}{\beta} \sum_j \frac{(2r - d_{ij}(t))(\mathbf{p}_i(t) - \mathbf{p}_j(t))}{d_{ij}(t)} - \mathbf{v}(t) \quad (7)$$

The three forces are then summed up together with a weighing factor (mass) to determine the acceleration.

$$\mathbf{a}(t) = \frac{1}{m}(\mathbf{f}_f(t) + \mathbf{f}_r(t) + \mathbf{f}_{i,r}(t)) \quad (8)$$

## 3. EXPERIMENTS ON PEDESTRIAN CROSSING

As a case study for the numerical experiments, policy analysis on a pedestrian crossing was done.

A typical pedestrian crossing is a "mix-lane" as shown in Figure 3 where pedestrians from both directions meet in the middle of the crossing and interact to avoid each other. As the results of those interactions, it may slow down the walking speed of the pedestrians and increase the delay and dissipation time to cross the road for the same number of pedestrians.

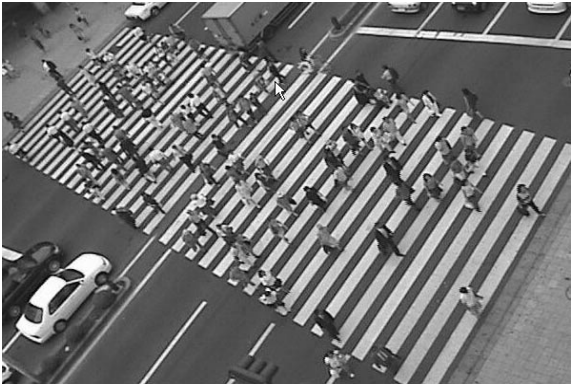

Figure 3. Typical pedestrian crossing

A simple policy such as to keep left (or right), or called "lane-like segregation" might be proposed to reduce the interaction. The implementation of this policy is straightforward. It can be done by marking an arrow on the left side of the starting point of the zebra cross. Using those markings, the pedestrians might be guided to keep left during the crossing. The reduction of the interaction due to lane-like segregation policy may increase the average walking-speed; reduce the delay and the dissipation time.

Using the numerical simulation model as explained in section 2, an experiment on pedestrian crossing was done in two scenarios. The existing condition is called "mix-lane" where pedestrians' initial and target locations are randomly generated at both ends of the crossing. The keep right policy or the "lane-like segregation" was implemented by generating the pedestrians in lower half (for west to east) and above half (for east to west). Figure 4 shows the condition of both scenarios before they meet in the middle of the crossing.

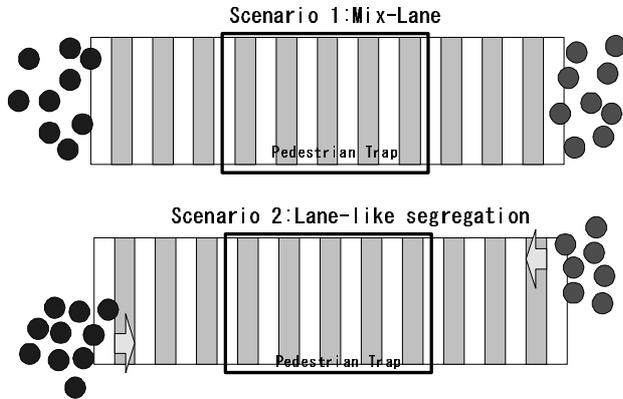

Figure 4. Two scenarios of the experiments of pedestrian crossing

Before the experiments were done, the simulation was calibrated using real pedestrian behavior. The maximum speed is set to be 1.2 m/s, the diameter of pedestrian is 70 cm and the sight distance is 4 meters. The parameter $\alpha$ was found to be 15, while parameter $\beta = 6$ and mass $m = 2$.

Some performance to represent the system and pedestrian behavior is determined. The simulation screen represents the pedestrian trap, which is in the middle of the crossing. The performance is only calculated when the pedestrians are in the trap. The instantaneous delay is the time difference between walking with the max speed and walking with the average speed. The dissipation time is calculated from the first pedestrian who enters the trap until the last pedestrian goes out of the trap.

Average performance of the system is calculated at each time. For example, if $\delta(t)$ is the average of instantaneous delay, that is sum of all instantaneous delays of all pedestrians divided by number of pedestrians in the trap, then the average delay of the system is calculated as

$$\bar{\delta}(t+1) = \frac{t-1}{t}\bar{\delta}(t) + \frac{\delta(t)}{t} \qquad (9)$$

The same formulation of average holds for average speed.

$$\bar{\mu}(t+1) = \frac{t-1}{t}\bar{\mu}(t) + \frac{\mu(t)}{t} \qquad (10)$$

## 4. RESULTS AND DISCUSSIONS

The experimental results in Figure 5 show that the increasing number of pedestrians will reduce the average speed almost linearly. The average speed of Lane-like segregation is higher than the mix lane. Higher number of pedestrians tends to increase the average speed difference between the two policies. Increasing the number of pedestrians in a lane-like segregation policy has a tendency to slower the drop of the average speed.

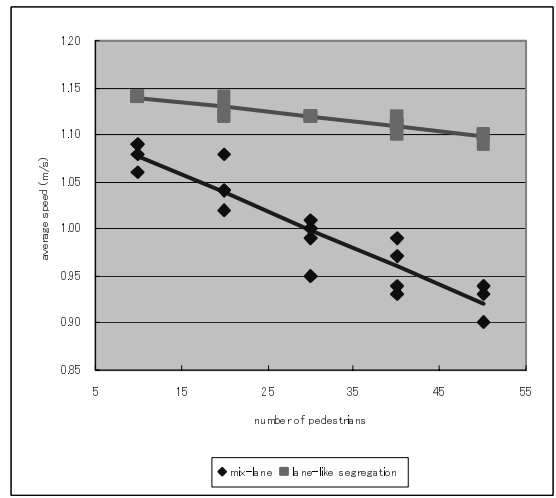

Figure 5. Relationship of average speed and number of pedestrians

Figure 6 illustrates that the time delay is increasing as the number of pedestrians increase. The lane-like segregation policy has much lower delay than the mix-lane policy. The change of delay due to the change of the number of pedestrians is also very low for the lane-like segregation policy.

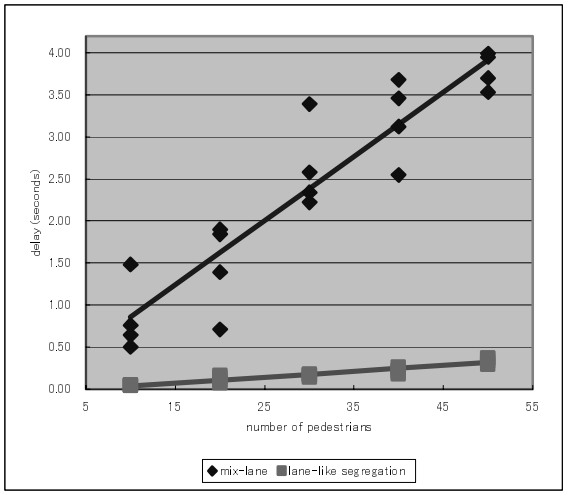

Figure 6. Relationship of delay and number of pedestrians

Figure 7 explains that the dissipation time is increasing as the number of pedestrian increase. Similar to the delay, the dissipation time of lane-like segregation is smaller than the mix-lane policy. It explains that for the same number of pedestrians that will cross the road, the lane-like segregation policy tend to reduce the time to cross due to less interaction.

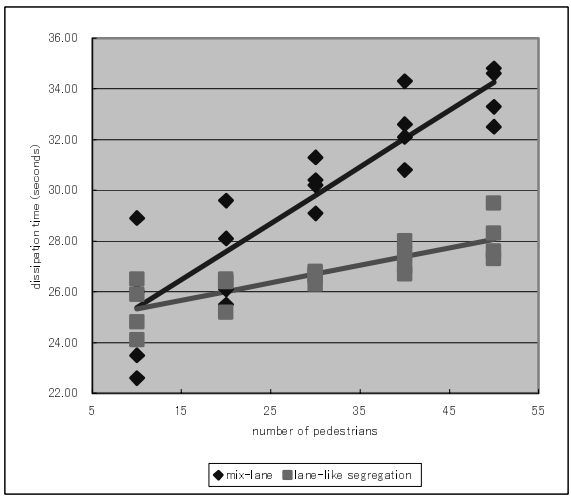

Figure 7. Relationship of dissipation time and number of pedestrians

## 5. CONCLUSIONS

To evaluate the impact of the policy quantitatively toward the behavior of pedestrians before its implementation, a microscopic pedestrian simulation model was developed. The model was based on physical forces, which work upon each pedestrian dynamically.

To demonstrate the numerical analysis of the model, an experimental policy on pedestrian crossing was performed. The simulation results showed that the keep-right policy or the lane-like segregation policy is inclined to be superior to do minimum or mix-lane policy in terms of average speed, average delay and dissipation time.

Validation of the simulation model toward the real world data is recommended for further study.

**REFERENCES**
1) Helbing, D. (1992) Models for pedestrian behavior. Natural Structures. Principles, Strategies, and Models in *Architecture and Nature, Part II*. Sonderforschungsbereich 230, Stuttgart, pp. 93-98.
2) Gipps, P.G. and Marksjo, B. (1985) A Micro-Simulation Model for Pedestrian Flows, *Mathematics and Computers in Simulation 27*, pp. 95-105.
3) Okazaki, S. (1979) A Study of Pedestrian Movement in Architectural Space, Part 1: Pedestrian Movement by the Application on of Magnetic Models, *Trans. of A.I.J.*, No.283, pp. 111-119.
4) Helbing, D. and Molnar, P. (1995) Social force model for pedestrian dynamics. *Physical Review E* 51, pp. 4282-4286.
5) Blue, V.J. and Adler, J.L. (2000) Cellular Automata Microsimulation of Bidirectional Pedestrian Flows, *Transportation Research Board* 1678, pp.135-141.
6) Lovas, G. G. (1994) Modeling and Simulation of Pedestrian Traffic Flow, *Transportation Research 28B*, pp. 429-443.
7) Thompson, P. A. and Marchant, E. W. (1995) A Computer Model the Evacuation of Large Building Populations, *Fire Safety Journal* 24, pp. 131-148.
8) Watts, J. M. (1987) Computer Models for Evacuation Analysis, *Fire Safety Journal* 12 pp. 237-245.